\documentclass[manuscript,screen]{acmart}  

\usepackage{graphicx}
\graphicspath{ {./images/} }
\usepackage{enumitem}
\usepackage{textcomp}
\usepackage{listings}
\usepackage{xcolor}
\usepackage{balance}

\definecolor{codegreen}{rgb}{0,0.5,0}
\definecolor{codegray}{rgb}{0.5,0.5,0.5}
\definecolor{codered}{rgb}{0.73,0.13,0.13}
\definecolor{backcolour}{rgb}{0.97,0.97,0.97}

\lstdefinestyle{mystyle}{
    backgroundcolor=\color{backcolour},   
    commentstyle=\color{codegreen},
    keywordstyle=\bf\color{codegreen},
    numberstyle=\tiny\color{gray},
    stringstyle=\color{codered},
    basicstyle=\ttfamily\footnotesize,
    breakatwhitespace=false,         
    breaklines=true,                 
    captionpos=b,                    
    keepspaces=true,                 
    numbers=left,                    
    numbersep=5pt,                  
    showspaces=false,                
    showstringspaces=false,
    showtabs=false,                  
    tabsize=2,
    frame=true,
    rulecolor=\color{backcolour},
}

\AtBeginDocument{%
  \providecommand\BibTeX{{%
    \normalfont B\kern-0.5em{\scshape i\kern-0.25em b}\kern-0.8em\TeX}}}

\copyrightyear{2023} 
\acmYear{2023} 
\setcopyright{none}\acmConference[SIGCSE 2023]{Proceedings of the 54th ACM Technical Symposium on Computer Science Education V. 1}{March 15--18, 2023}{Toronto, ON, Canada}
\acmBooktitle{Proceedings of the 54th ACM Technical Symposium on Computer Science Education V. 1 (SIGCSE 2023), March 15--18, 2023, Toronto, ON, Canada}
\acmPrice{15.00}
\acmDOI{10.1145/3545945.3569815}
\acmISBN{978-1-4503-9431-4/23/03}

\begin{document}

\title{GILP: An Interactive Tool for Visualizing the Simplex Algorithm}

\author{Henry W. Robbins}
\email{hwr26@cornell.edu}
\affiliation{%
  \institution{Cornell University}
  \city{Ithaca}
  \state{New York}
  \country{USA}
}

\author{Samuel C. Gutekunst}
\email{s.gutekunst@bucknell.edu}
\affiliation{%
  \institution{Bucknell University}
  \city{Lewisburg}
  \state{Pennsylvania}
  \country{USA}
}

\author{David B. Shmoys}
\email{david.shmoys@cornell.edu}
\affiliation{%
  \institution{Cornell University}
  \city{Ithaca}
  \state{New York}
  \country{USA}
}

\author{David P. Williamson}
\email{davidpwilliamson@cornell.edu}
\affiliation{%
  \institution{Cornell University}
  \city{Ithaca}
  \state{New York}
  \country{USA}
}

\begin{abstract}
The Simplex algorithm for solving linear programs---one of {\it Computing in Science \& Engineering}’s top 10 most influential algorithms of the 20th century---is an important topic in many algorithms courses. While the algorithm relies on intuitive geometric ideas, the computationally-involved mechanics of the algorithm can obfuscate a geometric understanding. In this paper, we present \texttt{gilp}, an easy-to-use Simplex algorithm visualization tool designed to connect the mechanical steps of the algorithm with their geometric interpretation. We provide an extensive library of example visualizations, and our tool allows instructors to quickly produce custom interactive HTML files for students to experiment with the algorithm (without requiring students to install anything!). The tool can also be used for interactive assignments in Jupyter notebooks, and has been incorporated into a forthcoming Data Science and Decision Making interactive textbook. In this paper, we first describe how the tool fits into the existing algorithm visualization literature: how it was designed to facilitate student engagement and instructor adoption, and how it substantially extends existing algorithm visualization tools for Simplex. We then describe the development and usage of the tool, and report feedback from its use in a course with roughly 100 students. Student feedback was overwhelmingly positive, with students finding the tool easy to use: it effectively helped them link the algebraic and geometrical views of the Simplex algorithm and understand its nuances. Finally, \texttt{gilp} is open-source, includes an extension to visualizing linear programming-based branch and bound, and is readily amenable to further extensions.
\end{abstract}



\maketitle

\section{Introduction} \label{sec:intro}
George Dantzig's Simplex algorithm \cite{dan51} for solving linear programs is an important topic in many algorithms courses.  The algorithm itself has been lauded as one of {\it Computing in Science \& Engineering}’s top 10 most influential algorithms of the 20th century \cite{don00} (see also \cite{nash00} for a more detailed discussion of the algorithm and its relevance), and it is one of the six algorithms from that list that also appeared in Nick Higham's 2016 list of The Top 10 Algorithms in Applied Mathematics \cite{hig16}.  A detailed treatment of linear programs and Simplex can be found in Chapter 29 of Cormen et al. \cite{cor09} (see also Levitin \cite{lev02}, where Simplex is described as the most important example of an iterative-improvement algorithm, and Kleinberg and Tardos \cite{kle06}, where Simplex is mentioned in the context of using linear programming to obtain a 2-approximation for the weighted Vertex Cover Problem).  Beyond algorithms courses, the Simplex algorithm and linear programming are fundamental concepts for many upper-division classes on optimization, operations research, and applied mathematics.

More formally, a \emph{linear program (LP)} is an optimization problem that can be expressed in the form: $\text{maximize } c^Tx$ subject to $Ax\leq b, x\geq 0,$ where $A\in \mathbb{R}^{m\times n}$, $c \in \mathbb{R}^n$, and $b\in \mathbb{R}^m$ are input matrices/vectors and we optimize $x$.  Often, LPs are motivated with a concrete example.  For example, Pendegraft \cite{pen97} and Cochran \cite{coc15} use an exercise where students are given a bag of Legos$^{\text{\textregistered}}$ with which they can build tables and chairs.  They are asked to do so in order to maximize profits, and are guided through formulating the problem mathematically as follows:

\begin{equation} \label{eq:lego}
\begin{split}
\mbox{maximize } & 16x_1 + 10x_2\\
\mbox{subject to: } \ & 2x_1+2x_2 \leq 8, \\
& 2x_1+x_2 \leq 6, \\
& x_1, x_2 \geq 0.
\end{split}
\end{equation}

\begin{figure}
    \includegraphics[width=0.4\paperwidth]{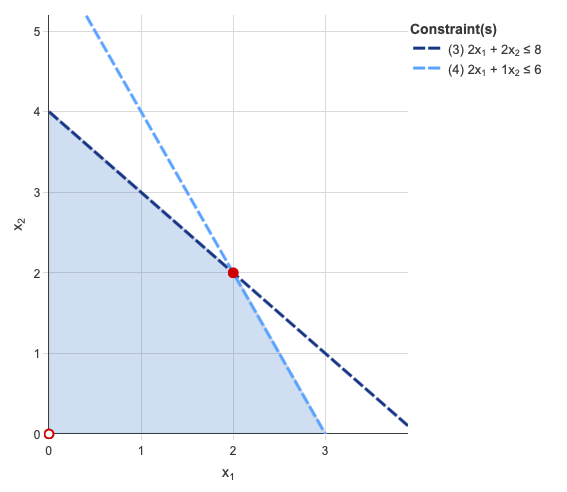}
    \caption{Geometry of an example linear program produced using \texttt{gilp}.}
    \label{fig:lego}
\end{figure}

Here, $x_1$ and $x_2$ respectively represent the number of tables and chairs they make, which can be sold respectively for \$16 and \$10 (leading to the \emph{objective function} $ \max 16x_1 + 10x_2$).  Each table and each chair requires 2 small Legos$^{\text{\textregistered}}$, and the bag they are given only contains 8 small Legos$^{\text{\textregistered}}$, which leads to the first constraint ($2x_1+2x_2 \leq 8$); the second constraint is similarly motivated.  To find an optimal solution, students first sketch the \emph{feasible region} (consisting of points $(x_1, x_2)$ satisfying all constraints); see Figure \ref{fig:lego}.  In this example, the optimal solution occurs at the \emph{corner point} at $x_1=x_2=2$, where the two constraint lines intersect.  In general, the feasible region of an LP is a polyhedron, and provided that the feasible region is nonempty and bounded, an optimal solution will always occur at one of these corner points.  See also, for example, Figure \ref{fig:klee_minty_3d}, which shows corner points for an LP with three variables. (More formally, these corner points are equivalently described as extreme points, vertices, and basic feasible solutions.  See Bertsimas and Tsitsiklis \cite{ber97} for details.)

\begin{figure*}
    \includegraphics[width=0.9\textwidth]{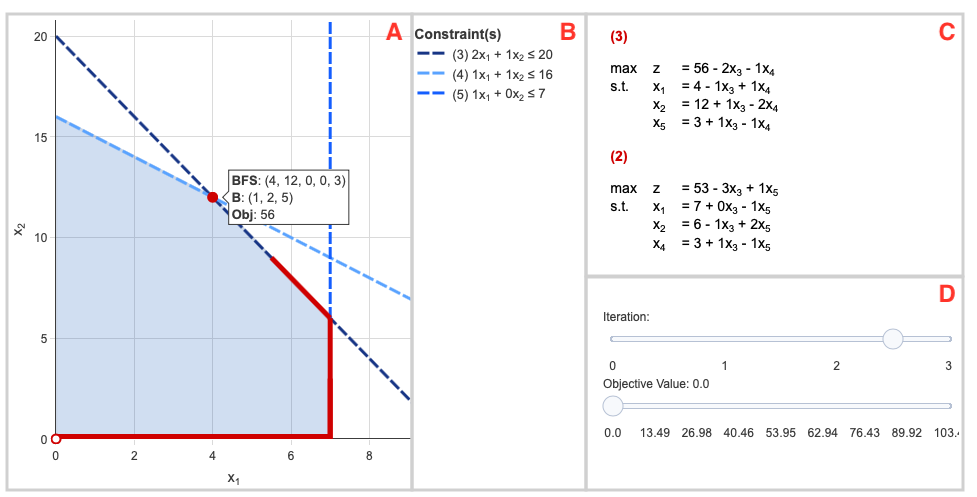}
    \caption{\texttt{gilp} Simplex visualization interface}
    \label{fig:simplex_visual_interface}
\end{figure*}

The Simplex algorithm finds an optimal solution to LPs following an intuitive geometric approach: start at some corner point, and move along the edges of the feasible region to better and better corner points, until finding a solution that is provably optimal (e.g., see the partial red path in Figure \ref{fig:simplex_visual_interface}).  Despite this geometric motivation, the algorithm is mechanically involved, and the algebraic manipulation can obfuscate the algorithm.  

\subsection{Simplex and LP Visualizations} \label{ssec:simp}

There are several visualizations for LPs, but as far as the authors are aware, no general tool that links the mechanical steps of the Simplex algorithm with what is happening geometrically.  Our main contribution is to provide such a tool which works in both two and three dimensions.  Our tool, \emph{Geometric Interpretation of Linear Programs (\texttt{gilp})}, was designed following best practices from Algorithm Visualization (AV) literature to facilitate both student engagement and instructor adoption.  

We begin, in Section \ref{sec:litrev}, by describing how our tool fits into AV literature.  We also discuss the novel features of our tool, relative to existing Simplex and LP visualization tools, which motivated the goals for our design.  Then, in Section \ref{sec:des}, we describe its design and use, emphasizing the flexibility it allows instructors (ranging from using pre-built examples, to incorporating it into highly-interactive Jupyter notebooks).  Finally, in Section \ref{sec:res}, we reflect on the tool's usage in a no-prerequisite course with roughly 100 students.

\section{Related Work and Motivation} \label{sec:litrev}
Algorithm Visualizations have a long history in the CS community, dating back to at least the 1981 \emph{Sorting out Sorting} video \cite{bae98}. By 2010, a survey paper analyzed a collection of over 500 AVs \cite{sha10}.  While a complete discussion of AV literature is beyond the scope of this paper, we begin by highlighting some pertinent results from the AV literature in Section \ref{ssec:AV}.  In Section \ref{ssec:simp} we then discuss  existing Simplex and LP visualizations, highlighting the novelty of \texttt{gilp}.

\subsection{Work on Algorithm Visualization} \label{ssec:AV}
There is an extensive literature on AV.  See, for example, Hundhausen, Douglas, and Stasko \cite{hun02}, Grissom, McNally, and Naps \cite{gri03}, Naps et al. \cite{naps02}, Urquiza-Fuentes and Vel{\'a}zquez-Iturbide \cite{urq09}, Fouh, Akbar, and Shaffer \cite{fouh12}, and Shaffer et al. \cite{sha10}.  

Pioneering work by Hundhausen et al. \cite{hun02} performed a meta-study on 24 AV experiments.  While this study showed mixed results on the effectiveness of AVs in general, it found that the most effective AVs  were those that were interactive and engaged students. 

The importance of interactive AVs has since been corroborated (see, e.g.,  Naps et al. \cite{naps02},  Grissom et al. \cite{gri03}, and Urquiza-Fuentes and Vel{\'a}zquez-Iturbide \cite{urq09}).  Naps et al. \cite{naps02} defines a taxonomy with six different forms of AV engagement (no viewing, viewing, responding  to questions about AVs, changing pieces of AVs, having students construct their own AVs, and presenting AVs), though notes that there is not a simple ordinal relationship between levels. 

AV literature has also pointed to a disconnect between instructor views of AVs and usage of AVs in classes.  Studies have routinely showed that CS instructors have an overwhelming belief in the efficacy of AVs.  Shaffer et al. \cite{sha11}, for instance, presents the results of a survey at SIGCSE 2010: 41 of 43 respondents either agreed or strongly agreed with the statement that ``Using algorithm visualizations can help learners learn computing science,'' with 0 respondents disagreeing or strongly disagreeing.  In contrast, only 22 of 41 respondents reported that they had used an AV in any class within the past two years.  See also Naps et al. \cite{naps02} and Isohanni and J{\"a}rvinen \cite{iso14}.  The reasons reported for this disconnect---for why many instructors don't use AVs despite believing in their efficacy---are fairly consistent.  Shaffer et al. \cite{sha11} reports that the two most common were: instructors had trouble finding suitable AVs, and instructors had trouble integrating AV material into courses; Naps et al. \cite{naps02} found that  major impediments to usage were related to instructor time (e.g., time developing or adapting AVs, time spent learning to use an AV, and time spent finding good examples).

One recent approach to designing AVs that are both easy for instructors to integrate into their classes and that involve student interaction has been to incorporate AVs into interactive textbooks.  See, for example, R{\"o}{\ss}ling et al. \cite{ros06}, Fouh, Akbar, and Shaffer \cite{fouh12}, and Smith et al. \cite{smi21}.  Smith et al. \cite{smi21} recently studied student use of a Jupyter notebook-based interactive textbook, and found that student's active engagement with the textbook and AVs were a significantly stronger predictor of student performance than traditional metrics like reading time.

{\bf Goal 1:} Our first goal was to design \texttt{gilp} based on best practices for facilitating both student engagement and instructor adoption (including the ability to readily embed \texttt{gilp} within Jupyter-notebook based interactive textbooks). 

\subsection{Simplex and LP Visualizations} \label{ssec:simp}

A typical introduction to LPs and the Simplex algorithm begins by showing how LPs with 2-3 variables can be solved using a {\bf graphical approach} (Figure \ref{fig:text} shows one such visualization).  While this technique does not work on LPs with more than three variables, it motivates the importance of corner points and thus the Simplex algorithm's geometric approach of moving along edges to better and better corner points.  Algebraically, the Simplex algorithm can be taught as systematically rewriting an LP in equivalent forms, until there is an ``obvious'' solution.  See, for instance, Section C of Figure \ref{fig:simplex_visual_interface}.  The two LPs (2) and (3) in that section have equivalent sets of feasible solutions (and both implicitly require that $x_i\geq 0$), but the LP (3), corresponding to the final step of the Simplex algorithm, has a solution that can readily be seen to be optimal: Set the right-hand-side variables $x_3$ and $x_4$ equal to zero, implying that $x_1=4, x_2=12$, and $x_5=3$.  This solution is feasible (since $x_i\geq 0$ and the equality constraints are met), has an objective function value of $z=56$, and is optimal as $x_3, x_4 \geq 0$ implies the objective function can never be larger than 56.  

It turns out that setting $x_3$ and $x_4$ (the right-hand-side variables) to 0 implies that there is no slack in the corresponding constraints listed in Section B of Figure \ref{fig:simplex_visual_interface}, indicating the corner point corresponding to this solution.  The visualization in  Figure \ref{fig:simplex_visual_interface} shows the user moving from the penultimate  LP (2) to the final equivalent LP (3), as the bold red line in Section A moves from the penultimate corner point to the final corner point (where the values of the $x_i$ are shown in the overlaid box which appears upon hovering).  The systematic rearrangement of these LPs---and deciding which variables enter and leave the right-hand-side---has a rich geometric interpretation, but involves several algebraic steps that can hinder a student's ability to understand the core geometric ideas.  For more details of Simplex, see, e.g., Chapter 29 of Cormen et al. \cite{cor09}.  

Over the past two decades, several different AVs have been developed to introduce core LP concepts (like the graphical approach) and help students perform the Simplex algorithm.  However, these tend to be focused either on just the algebraic steps of the Simplex algorithm or just on visualising LPs and the graphical approach that only works in two and three dimensions, and we are not aware of any AVs that connect the algebraic steps of the Simplex algorithm with their corresponding geometry on general examples.  Below, we survey the main LP and Simplex visualizations:
\begin{itemize}
\item Kydd \cite{kydd12} provides a web-based Java applet that helps students visualize 2-dimensional LPs and allows them to manipulate constraint lines and use the graphical approach. Kydd reports an experiment where students in sections of a class that used the applet performed significantly better than those who did not. However, this applet only illustrated LP basics in two dimensions, and does not include any Simplex visualization.  Moreover, the applet no longer appears accessible at the provided URL.  
\item Stanimirovi{\'c} et al. \cite{sta09} shows how to visualize two- and three-dimensional linear programs in Mathematica, focusing on visualizing the feasible region and using the graphical approach.  Fernandes and Pereira \cite{fer18} similarly use Mathematica to build GLP-Tool, a tool for visualizing two-dimensional LPs and the graphical approach.  Again, neither include any visualization of the Simplex algorithm.
\item Lazaridis et al. \cite{laz07} describes Visual LinProg, which shows the algebraic mechanics of the Simplex algorithm.  However, it does not appear to provide any corresponding geometric linkage, and it is no longer accessible at the provided URL.  Vanderbei's textbook Linear Programming \cite{van20} includes a companion webpage with online tools that help students with the mechanics of the Simplex algorithm, but these tools similarly don't show any geometric link.
\item Wojas and Krupa \cite{woj16} provides a visualization of how Simplex solves a single two-dimensional LP, showing the algebra and geometric interpretation of each step.  However, it is not interactive, no code is provided, and it consists of a single example.
\end{itemize}
Thus, while there are many existing AVs that help with LP visualization and the Simplex algorithm, no interactive tools connect the mechanics of the Simplex algorithm with the corresponding geometry. Similarly, many Simplex and LP visualizations are no longer accessible.

{\bf Goal 2:} Our second goal was to design the first AV connecting the mechanics of the Simplex algorithm with the corresponding geometry, in both two and three dimensions.

{\bf Goal 3:} Our final goal was to follow  best practices in open-source software and publish the tool in a public repository to ensure its longevity.

\section{Design \& Use} \label{sec:des}

Our tool, \texttt{gilp}, is a Python package for generating interactive visualizations of LPs and the Simplex algorithm. Figure \ref{fig:simplex_visual_interface} depicts the four main components of the interface (with grey borders and red labels added for reference). The feasible region and constraints defined by the linear program are shown in section A. By hovering over corner points, one can see the corresponding solution (including the values assigned to the \emph{slack variables} associated with each original constraint) and its objective value. By clicking constraints in section B, users can emphasize the line or plane defined by each constraint. Section D provides two sliders: iteration and objective value. The iteration slider moves through iterations of the Simplex algorithm. The LP corresponding to the current iteration is shown in section C. If the slider is toggled between two iterations, section C displays the LP at both iterations. By allowing the user to simultaneously interact with Simplex algebra and the corresponding geometrical interpretation, we achieve Goal 2. The objective value slider allows students to solve linear program using the graphical approach. 

Naps et al. \cite{naps2003} studies the educational impacts of visualizations and the impediments to widespread adoption. Based on their analysis, they offer advice for AV designers. To meet Goal 1, the design of \texttt{gilp} closely follows these recommendations (with special attention to ``design for flexability'' and ``map to existing teaching and learning resources''). In Section \ref{sec:flex}, we describe the three primary methods of interaction with \texttt{gilp}. In Section \ref{sec:adopt}, we highlight some of the built-in examples included with the tool and how \texttt{gilp} can be easily adapted to courses. Lastly, we emphasize the extensionality of \texttt{gilp} in Section \ref{sec:extend}, including an extension we have written allowing students to visualize linear-programming-based branch and bound for integer programs.

\subsection{Design for Flexibility} \label{sec:flex}
Part of Goal 1 was to design \texttt{gilp} so that instructors had several options in how they used it to engage students. To achieve this, \texttt{gilp} is designed to be used in a variety of ways; these range in setup time and student interaction, allowing instructors maximum flexibility. Here, we discuss three levels of instructor use.

\textbf{Pre-built visualizations.} The quickest way to get started with \texttt{gilp} is using the pre-built visualizations available at the supporting website. The website includes Simplex visualizations for example linear programs. We have created examples that cover common Simplex and LP topics, so that gilp can be readily incorporated in standard courses.  For instance, Cormen et al. \cite{cor09} begins by showing a sample LP in two-dimensions and uses the graphical approach to motivate the importance of corner points, walks through the mechanics of the Simplex algorithm, discusses how to find an initial corner point (e.g., if setting all $x_i$'s to 0 is not feasible), discusses the concept of degeneracy, and includes a note on a famous example of Klee et al. \cite{klee1972} relating to running time; our existing library includes visualizations for all of these concepts.   This level of usage is fully web-based, and enables students to quickly interact with Simplex examples while requiring no setup on the instructor or student's behalf.  Our website also includes several specific example LPs used in Cormen et al. \cite{cor09} so that instructors using that textbook can readily incorporate \texttt{gilp}.

\textbf{Instructor-generated examples.} Instructors with working Python environments (or those using a Python Jupyter notebook through, say, Google Colab) can easily generate visualizations of custom two- and three-dimensional linear programs. To do so, they need only install \texttt{gilp} with the \texttt{pip} package manager. They can then quickly define a linear program in terms of the input matrices/vectors described in Section \ref{sec:intro}. The visualization can be interacted with directly, or even exported to static HTML so that it can be embedded on a course website (where students can interact with them without installing anything). See Figure \ref{fig:lego_code} for example code\footnote{Future code snippets will omit \texttt{import} statements.}. 

\begin{figure}
\begin{lstlisting}[language=Python]
from gilp import LP
from gilp import simplex_visual

lp = LP(A=[[2,2],
           [2,1]],
        b=[8,6],
        c=[16,10])
visual = simplex_visual(lp=lp)

visual.write_html("lego.html")
\end{lstlisting}
\caption{Code to generate HTML for a Simplex visualization of LP (\ref{eq:lego}).}
\label{fig:lego_code}
\end{figure}

\begin{figure}
    \includegraphics[width=1\columnwidth]{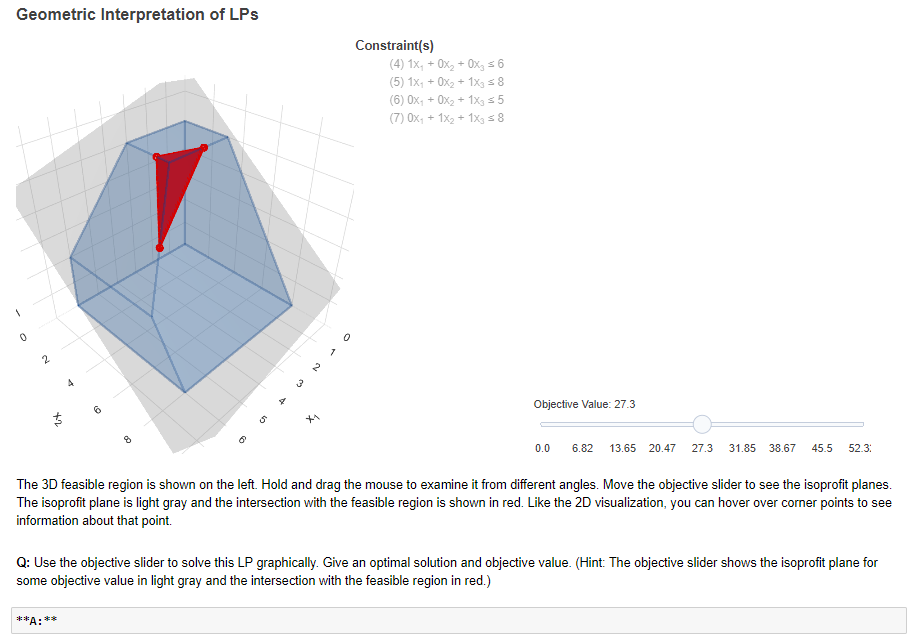}
    \caption{Student interface with \texttt{gilp} assignment in an interactive textbook.  }
    \label{fig:text}
\end{figure}

\textbf{Jupyter Notebook environment.} This method requires the heaviest lift as students must have a working Python environment with \texttt{jupyter} installed. However, this is an increasingly common expectation and tools like Google Colab provide students access to a Python environment through a web browser, without requiring students to download any software. Through the Jupyter Notebook interface, students can also alter and define their own linear programs. Furthermore, Jupyter Notebook's cell-based interface allows students to answer questions as they interact with the visualizations they create.  Figure \ref{fig:text} shows an example of this interface introducing students to the graphical approach in three dimensions. Students see text guiding them through the visualization and respond to questions. They can manipulate and rotate the visualization, hover over a corner point to see the corresponding solution, use the slider to highlight all solutions of a constant value (using the graphical approach to find the optimal solution and see the importance of corner points), and change the underlying LP.

Finally, and in line with Goal 1, we note that \verb|gilp| has been designed to allow usage at all levels of the taxonomy of Naps et al. \cite{naps02} described in Section \ref{ssec:AV}: students can view and interact with examples; integration into Jupyter notebook assignments and interactive textbooks enables students to respond to questions about visualizations and easily change/modify the visualization; the library of examples and ease with which students can create visualizations of custom LPs enables students to find and present examples highlighting important features of Simplex; as we will discuss in Section \ref{sec:extend}, a student with sufficient Python background can also extend \texttt{gilp} to create AVs for other related topics.

\subsection{Design for Instructor Adoption} \label{sec:adopt}
As part of Goal 1, we also wanted to make it as easy as possible for instructors to integrate \texttt{gilp} with existing teaching materials. For example, \texttt{gilp} is adaptable to different presentation styles of linear programming and the Simplex algorithm. The default representation of the linear program (i.e., what is shown in Section C of Figure \ref{fig:simplex_visual_interface}) is \emph{dictionary form} as presented by Chv\'{a}tal \cite{chvatal1983} and Cormen et al. \cite{cor09}.  However, many instructors use a different \emph{tableau form} (e.g. Levitin \cite{lev02}); \texttt{gilp} includes the options to display this format instead of dictionary form.

Additional configuration includes the ability to omit the value of slack variables from the feasible solution shown when hovering over a ``corner point'' and gives the instructor the choice of whether or not to show the corresponding dictionary or basis (allowing the instructor to also customize the visualization based on whether or not their audience knows linear algebra). These configurations are set with the \verb|basic_sol| and \verb|show_basis| parameters respectively.

\begin{figure}
\includegraphics[width=0.9\columnwidth]{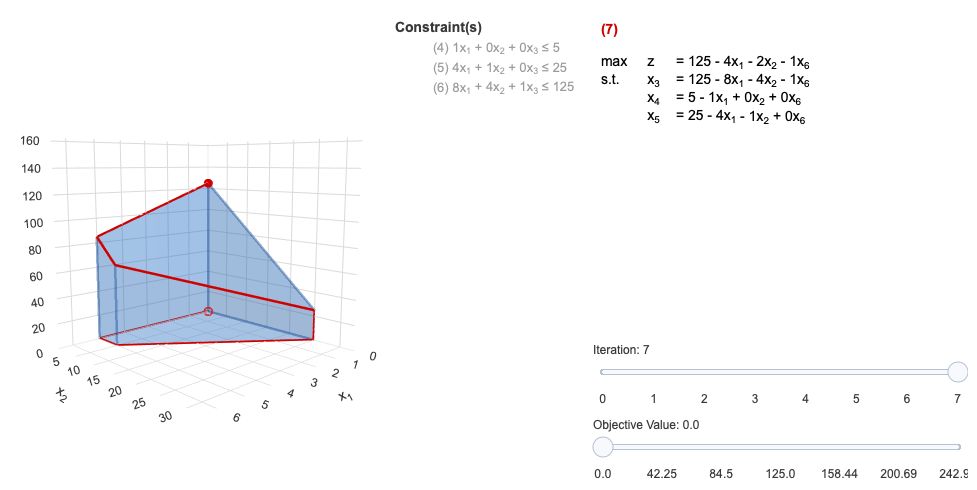}
\begin{lstlisting}[language=Python]
lp = examples.KLEE_MINTY_3D_LP
simplex_visual(lp=lp, rule="dantzig")
\end{lstlisting}
\caption{Simplex visualization of Klee-Minty cube.}
\label{fig:klee_minty_3d}
\end{figure}

In addition to providing numerous configuration options, \texttt{gilp} has multiple built-in examples to illustrate many concepts that instructors may wish to emphasize. For example, Klee et al. \cite{klee1972} demonstrate that the Simplex algorithm is not guaranteed to run in polynomial time through the construction of the Klee-Minty cube in variable dimension. In the $n$th dimension, the polytope has $2^n$ corner points. In the worst case, the path of Simplex will visit every corner point. Both the $n=2$ and $n=3$ case are included as built-in examples (see Figure \ref{fig:klee_minty_3d}). Other examples emphasize topics such as degeneracy and cycling, non-unique optimal solutions, and the integrality property.

\subsection{Branch and Bound \& Design for Extensionality} \label{sec:extend}

\begin{figure}
\includegraphics[width=0.9\columnwidth]{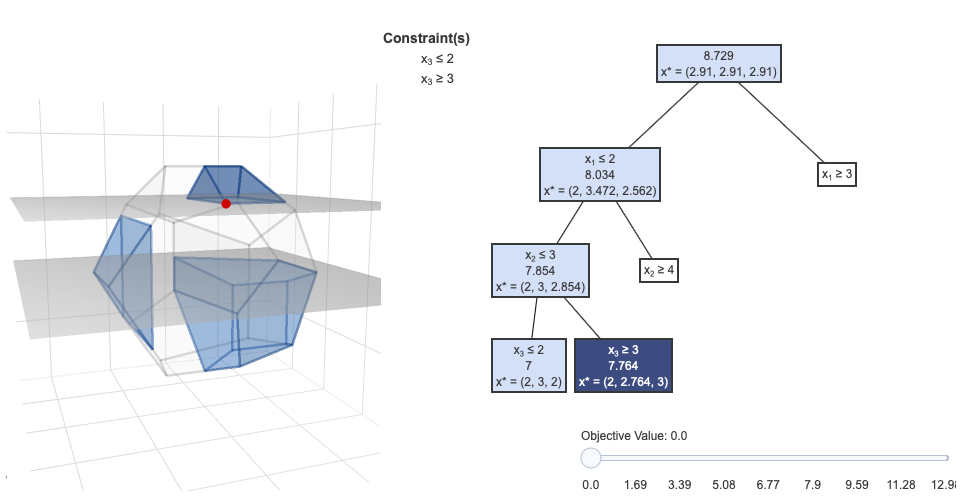}
\begin{lstlisting}[language=Python]
nodes = bnb_visual(lp=examples.DODECAHEDRON_3D_LP)
nodes[4]
\end{lstlisting}
\caption{Branch and bound visualization of a dodecahedron. }
\label{fig:dodecahedron}
\end{figure}

\texttt{gilp} is an actively maintained open-source project following best practices. As such, the tool is readily extendable to visualizing other linear and integer programming concepts such as cutting planes, interior-point methods, and sensitivity analysis. For example, visualizations of the branch and bound algorithm for solving integer programs are already available. The function \verb|bnb_visual| returns a list of figures, each representing a node explored in the branch and bound tree. For each figure, the node being explored is highlighted as well as the corresponding feasible region. The full path of Simplex is shown. Lastly, the two constraints generated from the parent node are included. Figure \ref{fig:dodecahedron} depicts the 5th node explored in running branch and bound on the \verb|DODECAHEDRON_3D_LP| example and the code used to generate the visualization.

\subsection{Implementation Details}

\texttt{gilp} is implemented in Python and is published on the Python Package Index (PyPI). It utilizes \texttt{plotly} as the core visualization library. The source code is licensed under CC BY-NC-SA 4.0\footnote{https://creativecommons.org/licenses/by-nc-sa/4.0/} and is available on GitHub. Through the GitHub Stars program, users can endorse high-quality projects. The \texttt{gilp} repository currently has 21 stars from users around the world, including users in Australia, Brazil, Germany, Ukraine, and the United States. Documentation can be found at the \texttt{gilp} website \cite{rob22}.  By publishing \texttt{gilp} to the Python Package Index and making it open-source and readily available on GitHub, we ensure its longevity and meet Goal 3.

\section{Usage in Class}\label{sec:res}
We first piloted \texttt{gilp} in a Fall 2020 course on Data Science and Decision Making, which covers discrete optimization problems (e.g., the shortest path problem, minimum spanning trees) and linear/integer programming.  For these topics, the emphasis is on both algorithms and modeling.  The class targets first-year students and has no mathematical or programming prerequisites.  In the course, students used \texttt{gilp} in assignments that were part of a forthcoming Data Science and Decision Making interactive textbook \cite{shm22}.

Students ran \texttt{gilp} through a Jupyter notebook, where they interacted with Simplex visualizations; questions guided them through this process, and students responded by typing answers into the notebook, which they then submitted digitally.  See Figure \ref{fig:text} for an example of how students interacted with the visualization. 
 Students worked on the assignment during two-hour recitations with live support from TAs.   Students first used \texttt{gilp} for LP visualization and Simplex in one recitation assignment, and used \texttt{gilp} again for visualizing branch and bound in a subsequent recitation assignment.

In the end-of-course survey, we included multiple numerical and free-response questions to evaluate \texttt{gilp}. These questions included:
\begin{itemize}
\item Q1: The python package, gilp, improved my understanding of the simplex algorithm.
\item Q2: Understanding the geometrical interpretation of a linear program helped me better reason about linear programs in general.
\item Q3: Understanding the geometrical interpretation of the simplex and branch [and] bound algorithms helped [me] understand the mechanics of these algorithms.
\item Q4: The python package, gilp, taught me the geometrical interpretation of LPs, simplex, and branch [and] bound.
\end{itemize}
 The survey was sent to all 85 students enrolled in the class, 77 of whom responded. Responses to these questions are summarized in Table \ref{tab:res}, and were  quite positive.
 
\begin{table}
\begin{tabular}{|l|l|l|l|l|l|l|} \hline
 & SD & D & N & A & SA & NA  \\ \hline
Q1 & 1 & 4 & 5 & 28 & 37 & 2   \\ \hline
Q2 & 0 & 1 & 7 & 29 & 38 & 2 \\ \hline
Q3 & 1 & 1 & 10 & 34 & 29 & 2 \\ \hline
Q4 & 0 & 2 & 13 & 42 & 18 & 2 \\ \hline
\end{tabular}
\caption{Numerical feedback to \texttt{gilp} survey questions.  SD = Strongly Disagree, D = Disagree, N = Neutral, A = Agree, SA = Strongly Agree, NA = No Answer.} \label{tab:res}
\end{table}

Students were also asked several free-response questions, including:
\begin{itemize}
\item What did you particularly {\it enjoy} about the gilp components of the lab?
\item What did you particularly {\it dislike} about the gilp components of the lab?
\item How could gilp be improved to better accomodate students?
\item Was there any misconception you had about linear programs, simplex, or branch [and] bound that your interaction with gilp visualizations corrected? If yes, please briefly describe the misconception that was addressed.
\end{itemize}
Responses to these questions were also overwhelmingly positive.  Students  reported that the \texttt{gilp} visualizations were easy to use and manipulate (e.g., resizing and rotating three-dimensional visualizations, keeping track of variables), and that the three-dimensional visualizations were especially useful (e.g., ``I suppose I could never picture 3d LPs before using gilp.'').  Perhaps most impressively, students were able to precisely identify and describe parts of the Simplex algorithm that were clarified by the \texttt{gilp} visualizations, and indicated ways in which seeing the geometry helped demystify algorithmic ideas. For example, one student wrote ``I thought LP's were solved the same way everytime, but turnsout [sic] there are different ways to go around the edges of a body, gilp made that clear.''

Students reported two themes in their constructive feedback, both of which related to technical implementation details.  First, some students ran into minor issues working with the pre-release of \verb|gilp|, which have all been resolved by subsequent official releases. Second, there were some browser-specific rendering issues with \texttt{plotly} in Safari; these have been resolved by a \texttt{plotly} update. In Spring 2022, we used an official release of \texttt{gilp} and students reported no technical issues. The same survey from Fall 2020 was emailed to these students a few months after the completion of the course, and while the response rates were far lower, they were again overwhelmingly positive.

\begin{acks}
We thank students and course staff for feedback during the development of \texttt{gilp}, EY for the funding supporting this effort, and the referees for their helpful comments. Special thanks to Thomas Smith for valuable feedback on the tool's visual accessibility.
\end{acks}

\bibliographystyle{ACM-Reference-Format}
\balance
\bibliography{main}

\end{document}